\documentclass[%
 reprint,
 superscriptaddress,
 bibnotes,
 amsmath,amssymb,
 aps,
 prl,
floatfix,
]{revtex4-2}

\usepackage{xr}

\usepackage{xcolor}
\usepackage[normalem]{ulem}

\usepackage{graphicx}
\usepackage{dcolumn}
\usepackage{bm}
\DeclareMathOperator{\erfc}{erfc}
\DeclareMathOperator{\sgn}{sgn}
\DeclareMathOperator{\erf}{erf}

\begin{document}

\title{Diffusion through Permeable Interfaces: Fundamental Equations and their Application to First-Passage and Local Time Statistics}

\author{Toby Kay}
 \email{toby.kay@bristol.ac.uk}
\affiliation{
Department of Engineering Mathematics, University of Bristol\\
Bristol, BS8 1UB, United Kingdom 
} 
\author{Luca Giuggioli}%
 \email{Luca.Giuggioli@bristol.ac.uk}
\affiliation{
Department of Engineering Mathematics, University of Bristol\\
Bristol, BS8 1UB, United Kingdom 
}
\affiliation{
Bristol Centre for Complexity Sciences, University of Bristol\\
Bristol, BS8 1UB, United Kingdom }

\date{\today}

\begin{abstract}
The diffusion equation is the primary tool to study the movement dynamics of a free Brownian particle, but when spatial heterogeneities in the form of permeable interfaces are present, no fundamental equation has been derived. Here we obtain such an equation from a microscopic description using a lattice random walk model. The sought after Fokker-Planck description and the corresponding backward Kolmogorov equation are employed to investigate first-passage and local time statistics and gain new insights. Among them a surprising phenomenon, in the case of a semi-bounded domain, is the appearance of a regime of dependence and independence on the location of the permeable barrier in the mean first-passage time. The new formalism is completely general: it allows to study the dynamics in the presence of multiple permeable barriers as well as reactive heterogeneities in bounded or unbounded domains and under the influence of external forces.
\end{abstract}

\maketitle


Random movement is ubiquitous, appearing in many physical, biological and social systems, and is traditionally modelled by diffusion in a homogeneous environment. But, in realistic systems the homogeneity of the environment is often interspersed by spatial heterogeneities that interfere significantly with diffusive transport. In many instances these heterogeneities are due to the presence of permeable interfaces, often referred to as semi or partially permeable barriers. They appear at microscopic scales in different porous media such as biological tissue \cite{callaghan1991diffraction,mitra1992diffusion,latour1994time,mair1999probing,sen2003time,friedman2008principles}, but also at larger scales when whole organisms interact with chemical or physical cues \cite{murphy2019one,tishkovskaya2021bayesian,murphy2021role}.

Cell biology is replete with examples of permeable structures whose function is to regulate the flux of biochemical substances between different spatial regions \cite{phillips2012physical}. In magnetic imaging techniques the diffusion of water molecules through different cellular compartments is exploited to understand physiological and anatomical properties of the human body \cite{grebenkov2014exploring1,grebenkov2014exploring2}. The lateral movement of molecules within the bilayer plasma membrane of eukaryotic cells is inhibited by the formation of submicron compartments due to anchored-transmembrane proteins and other macromolecules bound to the underlying actin-based cytoskeleton network \cite{kusumi2005paradigm}. Permeability is also of relevance to ecology where animal dispersal is affected by the heterogeneity of the landscape e.g. the type of habitat \cite{beyer2016you,kenkre2021theory} or the presence of roads and fences \cite{assis2019road}.

Various theoretical approaches to study diffusion through permeable interfaces have been proposed in the past: Green's functions in discrete \cite{kenkre1982exciton,kenkre2008molecular,kosztolowicz2001random} and continuous space \cite{powles1992exact,dudko2004diffusion,hahn2012heat,kay2022defect}, spectral decompositions \cite{hahn2012heat,moutal2019diffusion} and scattering techniques \cite{novikov2011random}. These techniques, whilst valuable, have been limited in their scope as they either demand spatial symmetries, e.g. analytical Green's functions, or employ a coarse-grained representation of the heterogeneities, e.g. effective medium approximations. In addition, these various approaches have failed to construct a unified framework capable of representing the diffusive dynamics with both permeable and reactive heterogeneities and to derive important quantities, such as first-passage and local time (or other Brownian functionals) statistics (i.e. through a backward Fokker-Planck representation). Given the wide-spread occurrence of permeable membranes, the above limitations call for the development of a fundamental theory of diffusion through permeable interfaces. 

In this letter we aim to provide such theory through a fully analytic treatment of the problem. Firstly we show how the permeable boundary condition arises from microscopic considerations in a simple unbiased lattice random walk model. Such model allows us to derive an inhomogeneous diffusion equation (DE), where the inhomogeneity accounts for the presence of a porous barrier. Extensions to the general case of finite domains and when an external force is present are also provided. As applications of our formalism we study explicitly first-passage and local time statistics of diffusion with a permeable barrier. 

{\it{Theoretical derivation}: }We consider a nearest-neighbour unbiased random walker on an infinite 1D lattice. The jump rate of the random walk between neighbouring sites equals $F$ except between the lattice points $r$ and $r+1$ where the rate is $f$ with $F>f$. The Master equation that represents the dynamics of the occupation probability, $\mathcal{P}_{m}(t)$, of the random walker at the $m$-th lattice point can be constructed as follows \cite{kenkre2008molecular}, 
\begin{multline}\label{eq:master}
    \frac{d \mathcal{P}_m(t)}{dt}=F[\mathcal{P}_{m+1}(t) + \mathcal{P}_{m-1}(t)-2 \mathcal{P}_m(t)] \\
    - \Delta[\mathcal{P}_{r+1}(t) - \mathcal{P}_r(t)](\delta_{m,r}- \delta_{m,r+1}),
\end{multline}
where $\Delta=F-f$ accounts for a partially reflecting defect between the sites $r$ and $r+1$ and $\delta_{m,r}$ is the Kronecker delta. With the help of the so-called defect technique \cite{kenkre2021memory} Eq. (\ref{eq:master}) is solved \footnote{See Supplemental Material at [URL will be inserted by publisher] for mathematical details and derivations.}. 

With the lattice spacing $\alpha \to 0$, we let $m,r,f,F$ become infinitely large such that $m\alpha\to x$, $r\alpha\to x_b$, $f\alpha \to \kappa$, $F\alpha^2 \to D$ and $\mathcal{P}_m(t)/\alpha \to P(x,t)$. That is $P(x,t)$ is the probability density for a diffusing particle (with a diffusion coefficient $D$) to be at the spatial position $x$ at time $t$ with a barrier located at $x_b$ whose permeability is $\kappa$, with units of velocity. One can show \cite{Note1} that $P(x,t)$ in this case satisfies the (DE), $\partial_t P(x,t)=\partial^2_{x}P(x,t)$, with the so-called leather or permeable boundary condition (PBC) \cite{tanner1978transient,powles1992exact,kosztolowicz2001membrane},
\begin{equation}\label{eq:permeable_bc}
    J(x_b^\pm,t)=\kappa[P(x_b^-,t)-P(x_b^+,t)].
\end{equation}
Here, the $\pm$ superscript denotes the respective side of the barrier and $J(x,t)=-D\partial_x P(x,t)$ is defined as the probability current. In other words we have proved that the continuum limit of $\mathcal{P}_m(t)$ becomes the solution of the DE with the PBC, Eq. (\ref{eq:permeable_bc}).

We now proceed to derive a more practical equation to study Brownian dynamics through permeable structures, by taking the continuum limit of Eq. (\ref{eq:master}). We utilise the relationship between the continuous limit of a finite difference and a derivative. In that limit, the left-hand side (LHS) and the first term on the right-hand side (RHS) of Eq. (\ref{eq:master}) corresponds to the DE. For the last term in Eq. (\ref{eq:master}) we consider the spatially discrete form of the probability current, $\mathcal{J}_m(t)=F[\mathcal{P}_m(t)-\mathcal{P}_{m+1}(t)]$, with $F$ replaced by $f$ when $m=r$ \footnote{$\mathcal{J}_r(t)=f[\mathcal{P}_r(t)-\mathcal{P}_{r+1}(t)]$ is the spatially discrete analogue of the PBC.} and we rewrite $\Delta[\mathcal{P}_{r+1}(t) - \mathcal{P}_r(t)](\delta_{m,r}- \delta_{m,r+1})$ as $(\Delta/f)\mathcal{J}_r(t)(\delta_{r+1,m}-\delta_{r,m})$. With the Kronecker delta becoming the Dirac delta function, we obtain the following inhomogeneous DE
\begin{equation}\label{eq:diffusion_permeable}
    \frac{\partial P(x,t)}{\partial t}=D \frac{\partial^2 P(x,t)}{\partial x^2} + \frac{D}{\kappa}\delta'(x-x_b)J(x_b,t),
\end{equation}
where $J(x,t)$ is the probability current as defined previously and $\delta'(x)$ represents the derivative of the Dirac delta function.   

Let us introduce the free propagator of the DE, $G_0(x,t|x_0)=\exp\left\{-(x-x_0)^2/4Dt \right\}/\sqrt{4 \pi Dt}$. The solution of Eq. (\ref{eq:diffusion_permeable}), with the localized initial condition $P(x,0)=\delta(x-x_0)$, is given in the Laplace domain (for any function $f(t)$, $\widetilde{f}(\epsilon)=\int_{0}^{\infty} f(t) e^{-\epsilon t} dt$) by \cite{Note1}
\begin{multline}\label{eq:smoluchowski_permeable_sol}
    \widetilde{P}(x,\epsilon|x_0)=\widetilde{G}_0(x,\epsilon|x_0)\\
    -\partial_{x_0}\widetilde{G}_0(x,\epsilon|x_b)\frac{\widetilde{J}_0(x_b,\epsilon|x_0)}{\frac{\kappa}{D}+ \partial_{x_0}\widetilde{J}_0(x_b,\epsilon|x_b)}.
\end{multline}
In Eq. (\ref{eq:smoluchowski_permeable_sol}) we have used the notation $P(x,t|x_0)$ to indicate the localized initial condition and $J_0(x,t|x_0)=-D \partial_x G_0(x,t|x_0)$ is defined as the free probability current ($\partial_{x} h(y)=\frac{\partial }{\partial x}h(x)|_{x=y}$ for a generic function $h(x)$) \footnote{{When the propagator is translationally invariant, one may replace $\partial_{x_0}\widetilde{G}_0(x,\epsilon|x_b)$ by $\widetilde{J}_0(x,\epsilon|x_b)/D$ and $\partial_{x_0}\widetilde{J}_0(x_b,\epsilon|x_b)$ by $-\partial_{x}\widetilde{J}_0(x_b,\epsilon|x_b)$}}. By inserting the correct propagator and its current into Eq. (\ref{eq:smoluchowski_permeable_sol}), one recovers the solution of the DE with the PBC (\ref{eq:permeable_bc}). 

\begin{figure}[htbp]
    \centering
    \includegraphics[width=8.6 cm]{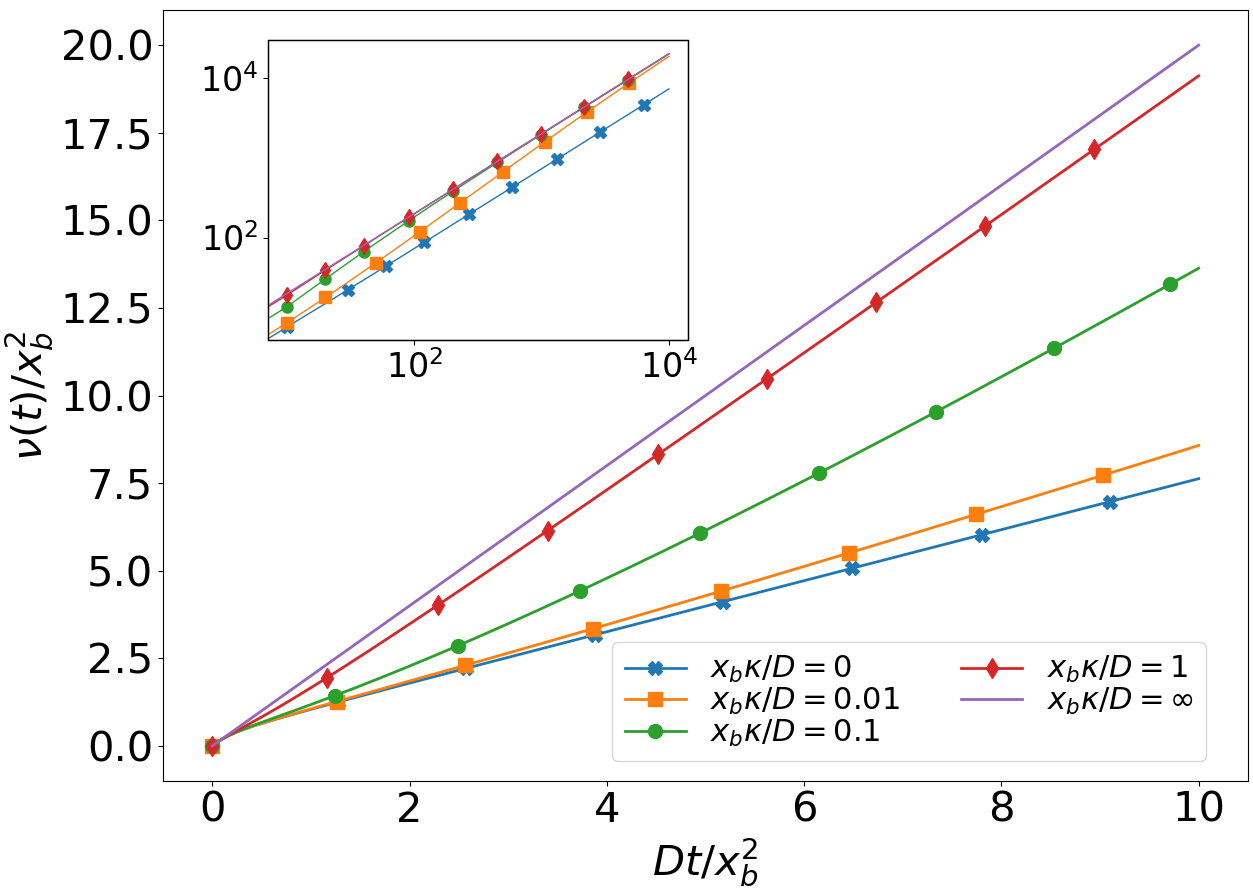}
    \caption{(Color online) The MSD, $\nu(t)$, as a function of time for a Brownian particle initially placed at the origin, in the presence of a permeable barrier, with permeability $\kappa$, placed at $x_b$, for different values of the scaled permeability parameter $x_b \kappa/D$. An infinite permeability indicates the absence of a barrier and a zero permeability indicates a fully reflecting barrier.
    Inset: corresponding long time behaviour of the MSD plotted against time on a logarithmic scale, showing how it has the asymptotic form $ 2Dt$, except for $\kappa=0$ which has $2(1-2/\pi)Dt$.}
    \label{fig:MSD}
\end{figure}

It is instructive to look at the moments of $P(x,t|x_0)$ i.e. $\langle x^n(t) \rangle = \int_{-\infty}^{\infty} x^n P(x,t|x_0) dx$. Using Eq. (\ref{eq:diffusion_permeable}) we find the following equations for the first and second moment, $\frac{d}{dt}\langle x(t) \rangle=-D J(x_b,t)/\kappa$ and $\frac{d}{dt}\langle x^2(t) \rangle=2D-2x_bD J(x_b,t)/\kappa$, respectively. As $J(x_b,t)$ is readily obtained from Eq. (\ref{eq:smoluchowski_permeable_sol}), these equations are solved by
\begin{equation}
    \langle x(t)\rangle=x_0-\sgn(x_b-x_0) \frac{D}{2\kappa}\beta(t)
\end{equation}
and
\begin{equation}\label{eq:first_moment}
    \langle x^2(t)\rangle=2Dt+x_0^2-\sgn(x_b-x_0) \frac{Dx_b}{\kappa}\beta(t),
\end{equation}
where $\sgn(z)$ is the sign function and
\begin{multline}\label{eq:second_moment}
    \beta(t)=\erfc\left\{ \frac{|x_0-x_b|}{2\sqrt{D t}}\right\} \\
    -\exp\left\{\frac{2\kappa}{D}\left(|x_0-x_b|+2 \kappa t \right) \right\} \erfc\left\{ \frac{|x_0-x_b|+4 \kappa t}{2\sqrt{D t}}\right\}.
\end{multline}
Here $\erfc(z)=1-\erf(z)$ with $\erf(z)$ the error function. In the limit of $\kappa\to \infty$ and $\kappa \to 0$ Eqs. (\ref{eq:first_moment}) and (\ref{eq:second_moment}) tend to their counterparts for free diffusion and diffusion with a perfectly reflecting boundary, respectively. As $\lim_{t\to\infty}\beta(t)=1$, the mean reaches a stationary value,  $\lim_{t\to\infty} \langle x(t) \rangle=x_0-\sgn(x_b-x_0)D/(2\kappa)$. In Fig. (\ref{fig:MSD}) we use Eqs. (\ref{eq:first_moment}) and (\ref{eq:second_moment}) to plot the mean square displacement (MSD) $\nu(t)=  \langle \left( x(t)-\langle x(t) \rangle \right)^2 \rangle$. The curves clearly show that the presence of the permeable barrier reduces the magnitude of the MSD for short times, yet at long times the $2Dt$ term is dominant and we have the standard diffusive linear increase.

We now rewrite Eq. (\ref{eq:diffusion_permeable}) in the following form, $\partial_t P(x,t)=L_xP(x,t)$, where $L_x$ is a linear differential operator with respect to $x$. To proceed, we exploit the property $\delta'(x-x_b)J(x_b,t)=\delta(x-x_b)\partial_x J(x,t)+\delta'(x-x_b)J(x,t)$, and the definition of $J(x,t)$, to write $L_x=(D^2/\kappa)\partial_x \delta'(x-x_b)+\partial^2_x[D-(D^2/\kappa)\delta(x-x_b)].$ The operator, $L$, corresponds to the one in the following Fokker-Planck equation (FPE) \cite{schuss2009theory}, 
\begin{equation}\label{eq:FP}
    \frac{\partial}{\partial t}P(x,t)=-\frac{\partial}{\partial x}\left[ A(x) P(x,t)\right]+\frac{\partial^2}{\partial x^2} \left[B(x) P(x,t)\right]
\end{equation}
with $A(x)=-(D^2/\kappa)\delta'(x-x_b)$ and $B(x)=D-(D^2/\kappa)\delta(x-x_b)$. Through this description we see that the presence of a permeable barrier can be described by an infinitely large positive potential, $(D^2/\kappa)\delta(x-x_b)$, that pushes away the Brownian particle from $x_b$, and by a diffusion coefficient that is modified at the interface, becoming infinitely negative thereby trapping the particle instead of dispersing it.

Though standard techniques allow to relate the underlying Langevin equation corresponding to Eq. (\ref{eq:FP}), the appearance of the Dirac delta function and its derivative would render such exercise of little practical use. Instead, we use the FPE to find the corresponding backward (Kolmogorov) FPE. In terms of $L$, the backward FPE is $-\partial_{t_0} P(x_0,t_0)=L_{x_0}^\dag P(x_0,t_0)$, where $L^\dag$ is the formal adjoint of $L$, i.e $L_{x_0}^\dag=A(x_0)\partial_{x_0}+B(x_0)\partial_{x_0}^2$. Note that this equation is now in terms of $x_0$ and $t_0$, where $t_0<t$. The adjoint is then, $L_{x_0}^\dag=-(D^2/\kappa)\delta'(x_0-x_b)\partial_{x_0}+[D-(D^2/\kappa)\delta(x_0-x_b)]\partial^2_{x_0}$, meaning $L$ is self-adjoint.

\begin{figure}[htbp]
    \centering
    \includegraphics[width=8.6 cm]{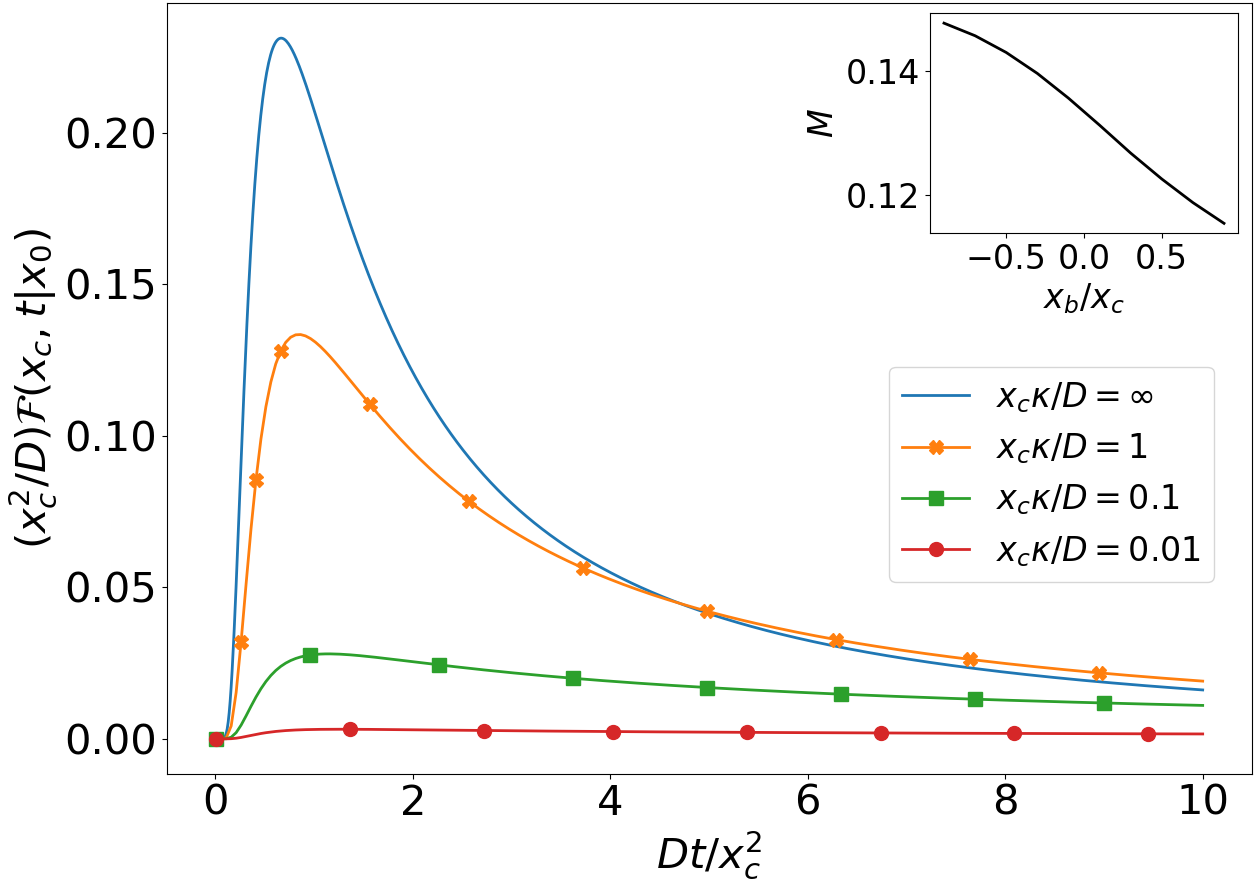}
    \caption{(Color online) The FPP distribution of a Brownian particle, $\mathcal{F}(x_c,t|x_0)$, is computed via numerical inversion \cite{abate2000introduction} of Eq. (\ref{eq:fpt}) for different values of the scaled permeability parameter, $x_c\kappa/D$. The particles starting location is $x_0$ with $x_0/x_c=-1$ and a permeable barrier is placed in between $x_0$ and $x_c$ at the origin. Inset: magnitude of the modal peak of the FPP distribution, $M$, plotted against different scaled barrier positions, $x_b/x_c$, with $x_0\kappa/D=-1$.}
    \label{fig:fpt}
\end{figure}

{\it{First-passage processes}: }Using the backward FPE we study the process in the presence of a perfectly absorbing point at $x_c$ to the left or right of both $x_0$ and $x_b$. Note, if this absorbing boundary is placed at the same point as the permeable barrier, $x_c=x_b$, a radiation boundary \cite{collins1949diffusion,waite1957theoretical,weiss1994aspects,redner2001guide} is recovered \cite{Note1}. Defining the survival probability as $S(t|x_0)=\int_{-\infty}^{x_c}P(x,t|x_0) dx$ or $S(t|x_0)=\int_{x_c}^{\infty}P(x,t|x_0) dx$, respectively, for $x_0<x_b<x_c$ or $x_c<x_b<x_0$. Taking $t_0=0$, exploiting the time homogeneity of the process and utilising the self-adjoint nature of $L$, we find that for $S(t|x_0)$, 
\begin{equation}\label{eq:survival}
    \frac{\partial S(t|x_0)}{\partial t}=D \frac{\partial^2 S(t|x_0)}{\partial x_0^2}-\frac{D^2}{\kappa}\delta'(x_0-x_b)\frac{\partial S(t|x_0)}{\partial x_0}\Big|_{x_0=x_b}.
\end{equation}
Eq. (\ref{eq:survival}) is supplemented by the initial condition, $S(0|x_0)=1$ and the Dirichlet boundary conditions (BC), $S(t|x_c)=0$ and $\lim_{x_0\to \pm\infty} S(t|x_0)=1$. Using the free propagator, we satisfy the Dirichlet BC via $G(x,t|x_0)=G_0(x,t|x_0)-G_0(x,t|2x_c-x_0)$ \cite{redner2001guide} and write the solution to Eq. (\ref{eq:survival}) as 
\begin{equation}\label{eq:survival_sol_laplace}
    \widetilde{S}(\epsilon|x_0)=\widetilde{S}_0(\epsilon|x_0)+\partial_{x_0} \widetilde{S}_0(\epsilon|x_b) \frac{\partial_x \widetilde{G}(x_b,\epsilon|x_0) }{\frac{\kappa}{D^2}-\partial^2_{x,x_0}\widetilde{G}(x_b,\epsilon|x_b)},
\end{equation}
where the free survival probability (i.e. for $\kappa=\infty$) is $S_0(t|x_0)=\erf \big\{|x_c-x_0|/\sqrt{4Dt}\big\}$ \cite{redner2001guide}. From $\widetilde{\mathcal{F}}(x_c,\epsilon|x_0)=1-\epsilon \widetilde{S}(\epsilon|x_0)$, we obtain the Laplace transform of the first-passage probability (FPP) distribution (see Ref. \cite{Note1} for the expression for when $x_0$ is between $x_b$ and $x_c$),
\begin{equation}\label{eq:fpt}
    \widetilde{\mathcal{F}}(x_c,\epsilon|x_0)=\frac{2\kappa e^{-|x_c-x_0|\sqrt{\frac{\epsilon}{D}}}}{\sqrt{D\epsilon}\left[1+e^{-2|x_c-x_b|\sqrt{\frac{\epsilon}{D}}}\right]+2\kappa}.
\end{equation}
Through Tauberian theorems \cite{feller2008introduction} we find the long time dependence of the FPP distribution as
\begin{equation}\label{eq:fpt_asymptotic}
    \mathcal{F}(x_c,t|x_0)\approx \frac{|x_c-x_0|+D/\kappa}{\sqrt{4 \pi D t^3}}.
\end{equation}
Eq. (\ref{eq:fpt_asymptotic}) shows that the FPP distribution possesses the same $t^{-3/2}$ asymptotic dependence as free diffusion but the coefficient includes the additional term $D/\kappa$. In Fig. (\ref{fig:fpt}) we draw Eq. (\ref{eq:fpt}) to show the full time dependence, while the inset shows the non-linear dependence of the magnitude of the mode of the distribution, $M$, as a function of the barrier position relative to $x_c$.

To gain further understanding of the impact a permeable barrier has on the dynamics of a Brownian particle, we study the mean first-passage time (MFPT) to $x_c$, $\tau(x_0)=\int_{0}^{\infty} t \mathcal{F}(x_c,t|x_0) dt$. Since, the MFPT of a Brownian particle is infinite in a semi-infinite domain, we add a perfectly reflecting boundary at $x_r$, such that the permeable barrier lies between $x_r$ and $x_c$. As $\tau(x_0)=\int_{0}^{\infty} S(t|x_0) dt$, from Eq. (\ref{eq:survival}) we have
\begin{equation}\label{eq:mfpt}
    -1=D\tau''(x_0)-\frac{D^2}{\kappa} \delta'(x_0-x_b)\tau'(x_b),
\end{equation}
where $\tau'(x_0)=\frac{d}{dx_0}\tau(x_0)$. Eq. (\ref{eq:mfpt}) is then supplemented by the Dirichlet and Neumann BC, $\tau(x_c)=0$ and $\tau'(x_r)=0$, respectively. Eq. (\ref{eq:mfpt}) is solved to give \cite{Note1},
\begin{equation}\label{eq:mfpt_sol}
    \tau(x_0)=\left\{
    \begin{array}{ll}
    \frac{x_c^2-x_0^2+2x_r(x_0-x_c)}{2 D}+\frac{|x_b-x_r|}{\kappa},\ x_0 \in [x_r,x_b),\\
    \frac{x_c^2-x_0^2+2x_r(x_0-x_c)}{2 D},\ x_0 \in (x_b,x_c].
    \end{array}
    \right .
\end{equation}
Eq. (\ref{eq:mfpt_sol}) shows the interesting feature that when the barrier is not placed between $x_0$ and $x_c$, the MFPT is identical to the barrier free case. Yet when the barrier is placed between $x_0$ and $x_c$ the impact to the MFPT is merely the addition of a term dependent on the position of the barrier that is scaled by the strength of its permeability. To clarify this aspect we may split the contributions to $\tau(x_0)$ between those trajectories that travel to $x_c$ without returning to $x_0$ and those that do return. The permeable interface clearly has no effect on the former trajectories as $x_b$ does not lie between $x_0$ and $x_c$. For the latter trajectories, the particle may return to $x_0$ multiple times before directly travelling to $x_c$ from $x_0$. Since the mean return time for an unbiased Brownian particle to any point is only dependent on the overall domain size \cite{kac1947notion}, the presence of a permeable interface will have no impact on these trajectories either.

\begin{figure}[htbp]
    \centering
    \includegraphics[width=8.6 cm]{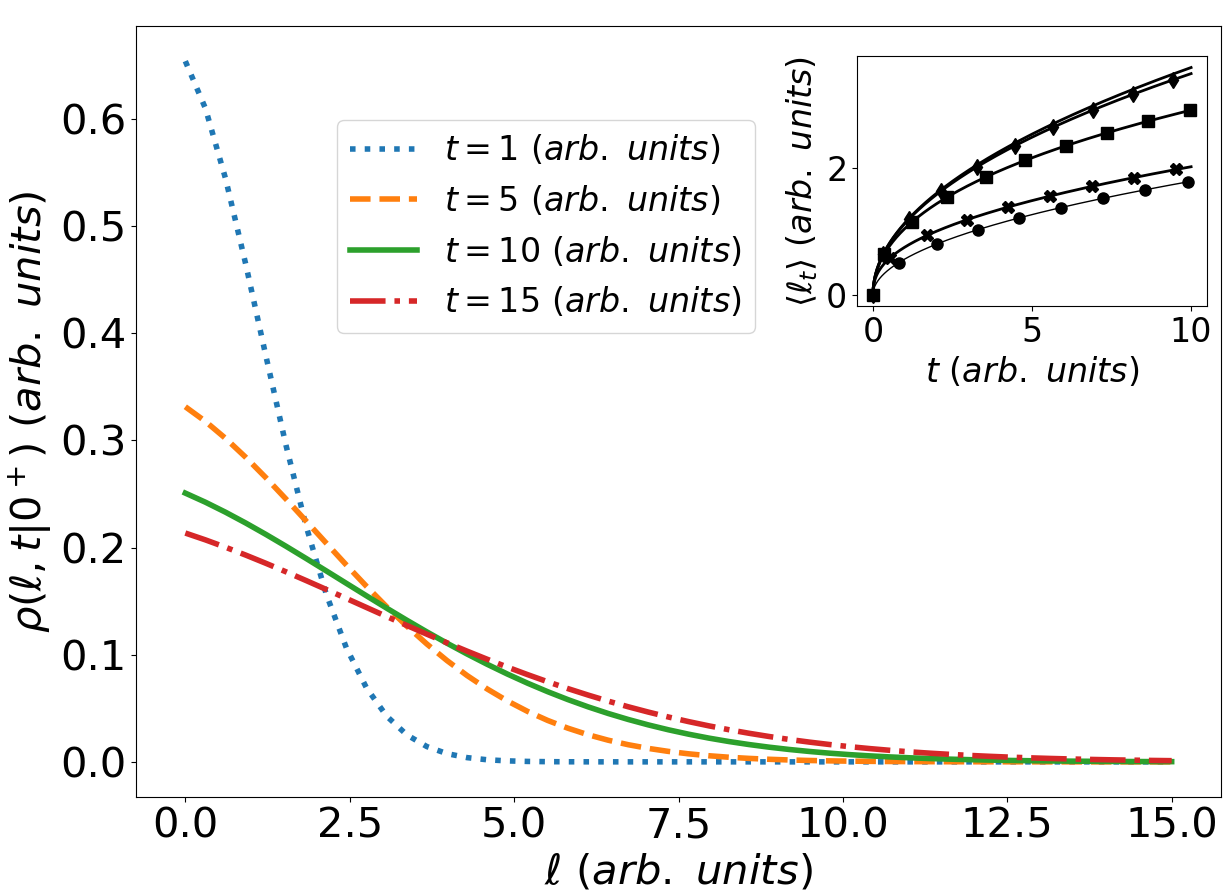}
    \caption{(Color online) The barrier local time distribution for $\kappa=0.1$ and $D=1$ (in arbitrary units), computed via a numerical inverse Laplace transform \cite{abate2000introduction} of Eq. (\ref{eq:local_time_dist}) and plotted against $\ell$ at different times, $t=1,5,10,15$, respectively.
    Inset: the mean barrier local time, Eq. (\ref{eq:local_time_mean}) plotted over a time window, for varying permeability values, $\kappa=\infty,1,0.1,0.01,0$ (in arbitrary units), represented by the markers: circular, cross, square, diamond and no marker, respectively. For $\kappa\to \infty$, we have the barrier free mean local time, $\sqrt{t/\pi D}$ and for $\kappa\to 0$, we have the perfectly reflecting barrier mean local time, $2\sqrt{t/\pi D}$.}
    \label{fig:local_time}
\end{figure}

{\it{Local time}: }Returning to the backward FPE, we can study the probability distribution of various functionals of Brownian motion. One of interest is the so-called local time of Brownian motion, defined as $\ell_t=\int_{0}^{t} \delta(x(t')-a)dt'$, which characterises the amount of time a Brownian particle spends at a given point $a$ \cite{levy1940certains}. We seek the probability density describing the random variable, $\ell_t$, namely the local time distribution (LTD), $\rho(\ell,t|x_0)$, of a Brownian particle in the presence of a permeable barrier. To do so we take the Laplace transform of the LTD with respect to $\ell$, i.e. $\varrho(p,t|x_0)=\int_{0}^{\infty}  \rho(\ell,t|x_0) e^{-p \ell} d\ell$. Such quantity may be written in terms of a conditional expectation \cite{majumdar2002local}, $\varrho(p,t|x_0)=\left\langle \exp\left\{-p\int_{0}^{t} \delta(x(t')-a)dt'\right\} \Big| x(0)=x_0 \right\rangle$, where the expectation is over all trajectories of the particle starting at $x(0)=x_0$ up to time $t$. Through the Feynman-Kac formula \cite{kac1949distributions,kac1951some}, $\varrho(p,t|x_0)$ satisfies the following
\begin{equation}\label{eq:Feynman-Kac}
    \frac{\partial \varrho}{\partial t}=A(x_0)\frac{\partial \varrho}{\partial x_0} + B(x_0)\frac{\partial^2 \varrho}{\partial x_0^2}-p\delta(x_0-a)\varrho,
\end{equation}
where $A$ and $B$ are defined after Eq. (\ref{eq:FP}). Eq. (\ref{eq:Feynman-Kac}) is supplemented by the initial condition, $\varrho(p,0|x_0)=1$, and the BC, $\varrho(p,t|x_0\to\pm\infty)=1$ \cite{majumdar2007brownian}. By treating the last term on the RHS of Eq. (\ref{eq:Feynman-Kac}) as an inhomogeneity, it is straightforward to construct the general solution via the solution of the homogeneous equation (i.e. for $p=0$). For a localized initial condition, the solution of the homogeneous part is equivalent to the solution of Eq. (\ref{eq:FP}) through Eq. (\ref{eq:smoluchowski_permeable_sol}). The Laplace transform of the solution of Eq. (\ref{eq:Feynman-Kac}) is thus
\begin{equation}\label{eq:FK_sol_lap}
    \widetilde{\varrho}(p,\epsilon|x_0)=\frac{1}{\epsilon}\left[1- \frac{\widetilde{P}(a,\epsilon|x_0)}{\frac{1}{p}+\widetilde{P}(a,\epsilon|a)}\right].
\end{equation}
Considering that we have a permeable barrier in an unbounded domain, we exploit the translational invariance of the problem and set $x_b=0$ and calculate the LTD at the barrier, that is $a=x_b$. Recalling the PBC (\ref{eq:permeable_bc}), we need to distinguish whether we are looking at $x_b^+$ or $x_b^-$. Furthermore, let us consider the case $x_b=0^+$ and $x_0=0^+$; using Eq. (\ref{eq:FK_sol_lap}) and after inverse Laplace transforming with respect to $p$, we find the barrier LTD to be 
\begin{equation}\label{eq:local_time_dist}
   \widetilde{\rho}(\ell,\epsilon|0^+)= \tfrac{\left(2 \kappa  \sqrt{D \epsilon }+D \epsilon \right) }{\epsilon  \left(\sqrt{D \epsilon }+\kappa \right)} \exp\left\{{-\tfrac{\left(2 \kappa  \sqrt{D \epsilon }+D \epsilon \right)}{\sqrt{D \epsilon }+\kappa }\ell}\right\}.
\end{equation}
The limit $\lim_{\epsilon\to 0}\epsilon \widetilde{\rho}(\ell,\epsilon|0^+)=0$ shows that Eq. (\ref{eq:local_time_dist}) has no steady state distribution at long times, indicative of the unbounded nature of the dynamics. In the limit $\kappa\to \infty$ we recover the barrier free distribution, $\rho(\ell,t|0)=2\sqrt{D/\pi t}e^{-D\ell^2/t}$ and for $\kappa\to 0$ we obtain the perfectly reflecting distribution, $\rho(\ell,t|0)=\sqrt{D/\pi t}e^{-D\ell^2/4t}$ \cite{takacs1995local}. From Eq. (\ref{eq:local_time_dist}) we can also find the mean,
\begin{equation}\label{eq:local_time_mean}
    \langle \ell_t \rangle = \frac{1}{4 \kappa} \left[ 1-e^{\frac{4 \kappa^2 t}{D}}\erfc\left\{2\kappa\sqrt{\tfrac{t}{D}}\right\} \right]+\sqrt{\tfrac{t}{\pi D}}.
\end{equation}
At long times the mean local time at the barrier is dominated by the final term on the RHS of Eq. (\ref{eq:local_time_mean}), i.e. $\langle \ell_t \rangle \sim t^{1/2}$, as in the barrier free case. A comparison of the temporal dependence of the mean local time, $\langle \ell_t \rangle$, for different values of permeability, is displayed in the inset of Fig. (\ref{fig:local_time}). The unbounded nature of its long time dependence can also be evinced from the main plot of Fig. (\ref{fig:local_time}), which shows the flattening of the LTD as time increases. 

{\it{External forces}: }We have shown so far the applications of our formalism to situations where no external forces are present. However the formalism is completely general and may include the dynamics in the presence of a potential, $U(x)$, in some domain $x \in \Omega$. In this case the `homogeneous' system is described by the Smoluchowski equation (SE) \cite{smoluchowski1915brownian}, $ \partial_t P(x,t)=\partial_x\left[U'(x) P(x,t)\right]+D \partial^2_x P(x,t)$. If we have a permeable barrier at $x_b\in \Omega$, the SE gets modified to
\begin{multline}\label{eq:smoluchowski_permeable}
    \frac{\partial P(x,t)}{\partial t}= \frac{\partial }{\partial x}[U'(x) P(x,t)]+D\frac{\partial^2 P(x,t)}{\partial x^2}\\
    +\frac{D}{\kappa}\delta'(x-x_b)J(x_b,t),
\end{multline}
where the probability current is now $J(x,t)=-U'(x)P(x,t)-D \partial_x P(x,t)$. Let us call the propagator of the SE, $G_0(x,t|x_0)$, which exists over $\Omega$, with $J_0(x,t|x_0)$ the barrier free counterpart of $J(x,t)$. The solution of Eq. (\ref{eq:smoluchowski_permeable}), with localized initial conditions, may be written as in Eq. (\ref{eq:smoluchowski_permeable_sol}). We are again able to transform Eq. (\ref{eq:smoluchowski_permeable}) into the FPE (\ref{eq:FP}), with $A(x)=-U'(x)[1-(D/\kappa)\delta(x-x_b)]-(D^2/\kappa)\delta'(x-x_b)$ and $B(x)=D-(D^2/\kappa)\delta(x-x_b)$, and then construct the analogous of Eqs. (\ref{eq:survival}), (\ref{eq:mfpt}) and (\ref{eq:Feynman-Kac}) in the presence of a potential. 

In summary, we have derived an inhomogeneous form of the DE and SE to account for the presence of a permeable barrier. We have used the former to investigate first-passage and local time statistics of a Brownian particle through the construction of a backward FPE. Explicit analytic dependence of the LTD and FPP distribution and their respective means have also been presented. Due to the linearity of the problem, our methods readily extend to the case of multiple permeable interfaces by appending the inhomogeneity for each interface position to Eq. (\ref{eq:smoluchowski_permeable}). Reactive heterogeneities can be accounted for in Eq. (\ref{eq:smoluchowski_permeable}) via the standard defect technique \cite{kenkre2021memory}. Future directions include the extension of these methodologies to higher dimensions and the application of our formalism to anomalous diffusion \cite{kosztolowicz2020boundary}.

\begin{acknowledgments}
TK and LG acknowledge funding from, respectively, the Engineering and Physical Sciences Research Council (EPSRC) Grant No. S100153-126 and the Biotechnology and Biological Sciences Research Council (BBSRC) Grant No. BB/T012196/1.
\end{acknowledgments}


\bibliography{bibliography}

\providecommand{\noopsort}[1]{}\providecommand{\singleletter}[1]{#1}%
\begin{thebibliography}{47}%
\makeatletter
\providecommand \@ifxundefined [1]{%
 \@ifx{#1\undefined}
}%
\providecommand \@ifnum [1]{%
 \ifnum #1\expandafter \@firstoftwo
 \else \expandafter \@secondoftwo
 \fi
}%
\providecommand \@ifx [1]{%
 \ifx #1\expandafter \@firstoftwo
 \else \expandafter \@secondoftwo
 \fi
}%
\providecommand \natexlab [1]{#1}%
\providecommand \enquote  [1]{``#1''}%
\providecommand \bibnamefont  [1]{#1}%
\providecommand \bibfnamefont [1]{#1}%
\providecommand \citenamefont [1]{#1}%
\providecommand \href@noop [0]{\@secondoftwo}%
\providecommand \href [0]{\begingroup \@sanitize@url \@href}%
\providecommand \@href[1]{\@@startlink{#1}\@@href}%
\providecommand \@@href[1]{\endgroup#1\@@endlink}%
\providecommand \@sanitize@url [0]{\catcode `\\12\catcode `\$12\catcode
  `\&12\catcode `\#12\catcode `\^12\catcode `\_12\catcode `\%12\relax}%
\providecommand \@@startlink[1]{}%
\providecommand \@@endlink[0]{}%
\providecommand \url  [0]{\begingroup\@sanitize@url \@url }%
\providecommand \@url [1]{\endgroup\@href {#1}{\urlprefix }}%
\providecommand \urlprefix  [0]{URL }%
\providecommand \Eprint [0]{\href }%
\providecommand \doibase [0]{https://doi.org/}%
\providecommand \selectlanguage [0]{\@gobble}%
\providecommand \bibinfo  [0]{\@secondoftwo}%
\providecommand \bibfield  [0]{\@secondoftwo}%
\providecommand \translation [1]{[#1]}%
\providecommand \BibitemOpen [0]{}%
\providecommand \bibitemStop [0]{}%
\providecommand \bibitemNoStop [0]{.\EOS\space}%
\providecommand \EOS [0]{\spacefactor3000\relax}%
\providecommand \BibitemShut  [1]{\csname bibitem#1\endcsname}%
\let\auto@bib@innerbib\@empty
\bibitem [{\citenamefont {Callaghan}\ \emph {et~al.}(1991)\citenamefont
  {Callaghan}, \citenamefont {Coy}, \citenamefont {MacGowan}, \citenamefont
  {Packer},\ and\ \citenamefont {Zelaya}}]{callaghan1991diffraction}%
  \BibitemOpen
  \bibfield  {author} {\bibinfo {author} {\bibfnamefont {P.~T.}\ \bibnamefont
  {Callaghan}}, \bibinfo {author} {\bibfnamefont {A.}~\bibnamefont {Coy}},
  \bibinfo {author} {\bibfnamefont {D.}~\bibnamefont {MacGowan}}, \bibinfo
  {author} {\bibfnamefont {K.~J.}\ \bibnamefont {Packer}},\ and\ \bibinfo
  {author} {\bibfnamefont {F.~O.}\ \bibnamefont {Zelaya}},\ }\bibfield  {title}
  {\bibinfo {title} {Diffraction-like effects in nmr diffusion studies of
  fluids in porous solids},\ }\href@noop {} {\bibfield  {journal} {\bibinfo
  {journal} {Nature}\ }\textbf {\bibinfo {volume} {351}},\ \bibinfo {pages}
  {467} (\bibinfo {year} {1991})}\BibitemShut {NoStop}%
\bibitem [{\citenamefont {Mitra}\ \emph {et~al.}(1992)\citenamefont {Mitra},
  \citenamefont {Sen}, \citenamefont {Schwartz},\ and\ \citenamefont
  {Le~Doussal}}]{mitra1992diffusion}%
  \BibitemOpen
  \bibfield  {author} {\bibinfo {author} {\bibfnamefont {P.~P.}\ \bibnamefont
  {Mitra}}, \bibinfo {author} {\bibfnamefont {P.~N.}\ \bibnamefont {Sen}},
  \bibinfo {author} {\bibfnamefont {L.~M.}\ \bibnamefont {Schwartz}},\ and\
  \bibinfo {author} {\bibfnamefont {P.}~\bibnamefont {Le~Doussal}},\ }\bibfield
   {title} {\bibinfo {title} {Diffusion propagator as a probe of the structure
  of porous media},\ }\href@noop {} {\bibfield  {journal} {\bibinfo  {journal}
  {Physical Review Letters}\ }\textbf {\bibinfo {volume} {68}},\ \bibinfo
  {pages} {3555} (\bibinfo {year} {1992})}\BibitemShut {NoStop}%
\bibitem [{\citenamefont {Latour}\ \emph {et~al.}(1994)\citenamefont {Latour},
  \citenamefont {Svoboda}, \citenamefont {Mitra},\ and\ \citenamefont
  {Sotak}}]{latour1994time}%
  \BibitemOpen
  \bibfield  {author} {\bibinfo {author} {\bibfnamefont {L.~L.}\ \bibnamefont
  {Latour}}, \bibinfo {author} {\bibfnamefont {K.}~\bibnamefont {Svoboda}},
  \bibinfo {author} {\bibfnamefont {P.~P.}\ \bibnamefont {Mitra}},\ and\
  \bibinfo {author} {\bibfnamefont {C.~H.}\ \bibnamefont {Sotak}},\ }\bibfield
  {title} {\bibinfo {title} {Time-dependent diffusion of water in a biological
  model system.},\ }\href@noop {} {\bibfield  {journal} {\bibinfo  {journal}
  {Proceedings of the National Academy of Sciences}\ }\textbf {\bibinfo
  {volume} {91}},\ \bibinfo {pages} {1229} (\bibinfo {year}
  {1994})}\BibitemShut {NoStop}%
\bibitem [{\citenamefont {Mair}\ \emph {et~al.}(1999)\citenamefont {Mair},
  \citenamefont {Wong}, \citenamefont {Hoffmann}, \citenamefont
  {H{\"u}rlimann}, \citenamefont {Patz}, \citenamefont {Schwartz},\ and\
  \citenamefont {Walsworth}}]{mair1999probing}%
  \BibitemOpen
  \bibfield  {author} {\bibinfo {author} {\bibfnamefont {R.~W.}\ \bibnamefont
  {Mair}}, \bibinfo {author} {\bibfnamefont {G.~P.}\ \bibnamefont {Wong}},
  \bibinfo {author} {\bibfnamefont {D.}~\bibnamefont {Hoffmann}}, \bibinfo
  {author} {\bibfnamefont {M.~D.}\ \bibnamefont {H{\"u}rlimann}}, \bibinfo
  {author} {\bibfnamefont {S.}~\bibnamefont {Patz}}, \bibinfo {author}
  {\bibfnamefont {L.~M.}\ \bibnamefont {Schwartz}},\ and\ \bibinfo {author}
  {\bibfnamefont {R.~L.}\ \bibnamefont {Walsworth}},\ }\bibfield  {title}
  {\bibinfo {title} {Probing porous media with gas diffusion nmr},\ }\href@noop
  {} {\bibfield  {journal} {\bibinfo  {journal} {Physical Review Letters}\
  }\textbf {\bibinfo {volume} {83}},\ \bibinfo {pages} {3324} (\bibinfo {year}
  {1999})}\BibitemShut {NoStop}%
\bibitem [{\citenamefont {Sen}(2003)}]{sen2003time}%
  \BibitemOpen
  \bibfield  {author} {\bibinfo {author} {\bibfnamefont {P.~N.}\ \bibnamefont
  {Sen}},\ }\bibfield  {title} {\bibinfo {title} {Time-dependent diffusion
  coefficient as a probe of the permeability of the pore wall},\ }\href@noop {}
  {\bibfield  {journal} {\bibinfo  {journal} {The Journal of Chemical Physics}\
  }\textbf {\bibinfo {volume} {119}},\ \bibinfo {pages} {9871} (\bibinfo {year}
  {2003})}\BibitemShut {NoStop}%
\bibitem [{\citenamefont {Friedman}(2008)}]{friedman2008principles}%
  \BibitemOpen
  \bibfield  {author} {\bibinfo {author} {\bibfnamefont {M.~H.}\ \bibnamefont
  {Friedman}},\ }\href@noop {} {\emph {\bibinfo {title} {Principles and Models
  of Biological Transport}}}\ (\bibinfo  {publisher} {Springer Science \&
  Business Media},\ \bibinfo {year} {2008})\BibitemShut {NoStop}%
\bibitem [{\citenamefont {Murphy}\ \emph {et~al.}(2019)\citenamefont {Murphy},
  \citenamefont {Buenzli}, \citenamefont {Baker},\ and\ \citenamefont
  {Simpson}}]{murphy2019one}%
  \BibitemOpen
  \bibfield  {author} {\bibinfo {author} {\bibfnamefont {R.~J.}\ \bibnamefont
  {Murphy}}, \bibinfo {author} {\bibfnamefont {P.~R.}\ \bibnamefont {Buenzli}},
  \bibinfo {author} {\bibfnamefont {R.}~\bibnamefont {Baker}},\ and\ \bibinfo
  {author} {\bibfnamefont {M.~J.}\ \bibnamefont {Simpson}},\ }\bibfield
  {title} {\bibinfo {title} {A one-dimensional individual-based mechanical
  model of cell movement in heterogeneous tissues and its coarse-grained
  approximation},\ }\href@noop {} {\bibfield  {journal} {\bibinfo  {journal}
  {Proceedings of the Royal Society A}\ }\textbf {\bibinfo {volume} {475}},\
  \bibinfo {pages} {20180838} (\bibinfo {year} {2019})}\BibitemShut {NoStop}%
\bibitem [{\citenamefont {Tishkovskaya}\ and\ \citenamefont
  {Blackwell}(2021)}]{tishkovskaya2021bayesian}%
  \BibitemOpen
  \bibfield  {author} {\bibinfo {author} {\bibfnamefont {S.~V.}\ \bibnamefont
  {Tishkovskaya}}\ and\ \bibinfo {author} {\bibfnamefont {P.~G.}\ \bibnamefont
  {Blackwell}},\ }\bibfield  {title} {\bibinfo {title} {Bayesian estimation of
  heterogeneous environments from animal movement data},\ }\href@noop {}
  {\bibfield  {journal} {\bibinfo  {journal} {Environmetrics}\ ,\ \bibinfo
  {pages} {e2679}} (\bibinfo {year} {2021})}\BibitemShut {NoStop}%
\bibitem [{\citenamefont {Murphy}\ \emph {et~al.}(2021)\citenamefont {Murphy},
  \citenamefont {Buenzli}, \citenamefont {Tambyah}, \citenamefont {Thompson},
  \citenamefont {Hugo}, \citenamefont {Baker},\ and\ \citenamefont
  {Simpson}}]{murphy2021role}%
  \BibitemOpen
  \bibfield  {author} {\bibinfo {author} {\bibfnamefont {R.~J.}\ \bibnamefont
  {Murphy}}, \bibinfo {author} {\bibfnamefont {P.~R.}\ \bibnamefont {Buenzli}},
  \bibinfo {author} {\bibfnamefont {T.~A.}\ \bibnamefont {Tambyah}}, \bibinfo
  {author} {\bibfnamefont {E.~W.}\ \bibnamefont {Thompson}}, \bibinfo {author}
  {\bibfnamefont {H.~J.}\ \bibnamefont {Hugo}}, \bibinfo {author}
  {\bibfnamefont {R.~E.}\ \bibnamefont {Baker}},\ and\ \bibinfo {author}
  {\bibfnamefont {M.~J.}\ \bibnamefont {Simpson}},\ }\bibfield  {title}
  {\bibinfo {title} {The role of mechanical interactions in emt},\ }\href@noop
  {} {\bibfield  {journal} {\bibinfo  {journal} {Physical Biology}\ }\textbf
  {\bibinfo {volume} {18}},\ \bibinfo {pages} {046001} (\bibinfo {year}
  {2021})}\BibitemShut {NoStop}%
\bibitem [{\citenamefont {Phillips}\ \emph {et~al.}(2012)\citenamefont
  {Phillips}, \citenamefont {Kondev}, \citenamefont {Theriot}, \citenamefont
  {Garcia},\ and\ \citenamefont {Orme}}]{phillips2012physical}%
  \BibitemOpen
  \bibfield  {author} {\bibinfo {author} {\bibfnamefont {R.}~\bibnamefont
  {Phillips}}, \bibinfo {author} {\bibfnamefont {J.}~\bibnamefont {Kondev}},
  \bibinfo {author} {\bibfnamefont {J.}~\bibnamefont {Theriot}}, \bibinfo
  {author} {\bibfnamefont {H.~G.}\ \bibnamefont {Garcia}},\ and\ \bibinfo
  {author} {\bibfnamefont {N.}~\bibnamefont {Orme}},\ }\href@noop {} {\emph
  {\bibinfo {title} {Physical Biology of the Cell}}}\ (\bibinfo  {publisher}
  {Garland Science},\ \bibinfo {year} {2012})\BibitemShut {NoStop}%
\bibitem [{\citenamefont {Grebenkov}\ \emph {et~al.}(2014)\citenamefont
  {Grebenkov}, \citenamefont {Van~Nguyen},\ and\ \citenamefont
  {Li}}]{grebenkov2014exploring1}%
  \BibitemOpen
  \bibfield  {author} {\bibinfo {author} {\bibfnamefont {D.~S.}\ \bibnamefont
  {Grebenkov}}, \bibinfo {author} {\bibfnamefont {D.}~\bibnamefont
  {Van~Nguyen}},\ and\ \bibinfo {author} {\bibfnamefont {J.-R.}\ \bibnamefont
  {Li}},\ }\bibfield  {title} {\bibinfo {title} {Exploring diffusion across
  permeable barriers at high gradients. i. narrow pulse approximation},\
  }\href@noop {} {\bibfield  {journal} {\bibinfo  {journal} {Journal of
  Magnetic Resonance}\ }\textbf {\bibinfo {volume} {248}},\ \bibinfo {pages}
  {153} (\bibinfo {year} {2014})}\BibitemShut {NoStop}%
\bibitem [{\citenamefont {Grebenkov}(2014)}]{grebenkov2014exploring2}%
  \BibitemOpen
  \bibfield  {author} {\bibinfo {author} {\bibfnamefont {D.~S.}\ \bibnamefont
  {Grebenkov}},\ }\bibfield  {title} {\bibinfo {title} {Exploring diffusion
  across permeable barriers at high gradients. ii. localization regime},\
  }\href@noop {} {\bibfield  {journal} {\bibinfo  {journal} {Journal of
  Magnetic Resonance}\ }\textbf {\bibinfo {volume} {248}},\ \bibinfo {pages}
  {164} (\bibinfo {year} {2014})}\BibitemShut {NoStop}%
\bibitem [{\citenamefont {Kusumi}\ \emph {et~al.}(2005)\citenamefont {Kusumi},
  \citenamefont {Nakada}, \citenamefont {Ritchie}, \citenamefont {Murase},
  \citenamefont {Suzuki}, \citenamefont {Murakoshi}, \citenamefont {Kasai},
  \citenamefont {Kondo},\ and\ \citenamefont {Fujiwara}}]{kusumi2005paradigm}%
  \BibitemOpen
  \bibfield  {author} {\bibinfo {author} {\bibfnamefont {A.}~\bibnamefont
  {Kusumi}}, \bibinfo {author} {\bibfnamefont {C.}~\bibnamefont {Nakada}},
  \bibinfo {author} {\bibfnamefont {K.}~\bibnamefont {Ritchie}}, \bibinfo
  {author} {\bibfnamefont {K.}~\bibnamefont {Murase}}, \bibinfo {author}
  {\bibfnamefont {K.}~\bibnamefont {Suzuki}}, \bibinfo {author} {\bibfnamefont
  {H.}~\bibnamefont {Murakoshi}}, \bibinfo {author} {\bibfnamefont {R.~S.}\
  \bibnamefont {Kasai}}, \bibinfo {author} {\bibfnamefont {J.}~\bibnamefont
  {Kondo}},\ and\ \bibinfo {author} {\bibfnamefont {T.}~\bibnamefont
  {Fujiwara}},\ }\bibfield  {title} {\bibinfo {title} {Paradigm shift of the
  plasma membrane concept from the two-dimensional continuum fluid to the
  partitioned fluid: high-speed single-molecule tracking of membrane
  molecules},\ }\href@noop {} {\bibfield  {journal} {\bibinfo  {journal} {Annu.
  Rev. Biophys. Biomol. Struct.}\ }\textbf {\bibinfo {volume} {34}},\ \bibinfo
  {pages} {351} (\bibinfo {year} {2005})}\BibitemShut {NoStop}%
\bibitem [{\citenamefont {Beyer}\ \emph {et~al.}(2016)\citenamefont {Beyer},
  \citenamefont {Gurarie}, \citenamefont {B{\"o}rger}, \citenamefont
  {Panzacchi}, \citenamefont {Basille}, \citenamefont {Herfindal},
  \citenamefont {Van~Moorter}, \citenamefont {R.~Lele},\ and\ \citenamefont
  {Matthiopoulos}}]{beyer2016you}%
  \BibitemOpen
  \bibfield  {author} {\bibinfo {author} {\bibfnamefont {H.~L.}\ \bibnamefont
  {Beyer}}, \bibinfo {author} {\bibfnamefont {E.}~\bibnamefont {Gurarie}},
  \bibinfo {author} {\bibfnamefont {L.}~\bibnamefont {B{\"o}rger}}, \bibinfo
  {author} {\bibfnamefont {M.}~\bibnamefont {Panzacchi}}, \bibinfo {author}
  {\bibfnamefont {M.}~\bibnamefont {Basille}}, \bibinfo {author} {\bibfnamefont
  {I.}~\bibnamefont {Herfindal}}, \bibinfo {author} {\bibfnamefont
  {B.}~\bibnamefont {Van~Moorter}}, \bibinfo {author} {\bibfnamefont
  {S.}~\bibnamefont {R.~Lele}},\ and\ \bibinfo {author} {\bibfnamefont
  {J.}~\bibnamefont {Matthiopoulos}},\ }\bibfield  {title} {\bibinfo {title}
  {‘you shall not pass!’: quantifying barrier permeability and proximity
  avoidance by animals},\ }\href@noop {} {\bibfield  {journal} {\bibinfo
  {journal} {Journal of Animal Ecology}\ }\textbf {\bibinfo {volume} {85}},\
  \bibinfo {pages} {43} (\bibinfo {year} {2016})}\BibitemShut {NoStop}%
\bibitem [{\citenamefont {Kenkre}\ and\ \citenamefont
  {Giuggioli}(2021)}]{kenkre2021theory}%
  \BibitemOpen
  \bibfield  {author} {\bibinfo {author} {\bibfnamefont {V.~M.}\ \bibnamefont
  {Kenkre}}\ and\ \bibinfo {author} {\bibfnamefont {L.}~\bibnamefont
  {Giuggioli}},\ }\href@noop {} {\emph {\bibinfo {title} {Theory of the Spread
  of Epidemics and Movement Ecology of Animals: An Interdisciplinary Approach
  Using Methodologies of Physics and Mathematics}}}\ (\bibinfo  {publisher}
  {Cambridge University Press},\ \bibinfo {year} {2021})\BibitemShut {NoStop}%
\bibitem [{\citenamefont {Assis}\ \emph {et~al.}(2019)\citenamefont {Assis},
  \citenamefont {Giacomini},\ and\ \citenamefont {Ribeiro}}]{assis2019road}%
  \BibitemOpen
  \bibfield  {author} {\bibinfo {author} {\bibfnamefont {J.~C.}\ \bibnamefont
  {Assis}}, \bibinfo {author} {\bibfnamefont {H.~C.}\ \bibnamefont
  {Giacomini}},\ and\ \bibinfo {author} {\bibfnamefont {M.~C.}\ \bibnamefont
  {Ribeiro}},\ }\bibfield  {title} {\bibinfo {title} {Road permeability index:
  evaluating the heterogeneous permeability of roads for wildlife crossing},\
  }\href@noop {} {\bibfield  {journal} {\bibinfo  {journal} {Ecological
  Indicators}\ }\textbf {\bibinfo {volume} {99}},\ \bibinfo {pages} {365}
  (\bibinfo {year} {2019})}\BibitemShut {NoStop}%
\bibitem [{\citenamefont {Kenkre}(1982)}]{kenkre1982exciton}%
  \BibitemOpen
  \bibfield  {author} {\bibinfo {author} {\bibfnamefont {V.~M.}\ \bibnamefont
  {Kenkre}},\ }\href@noop {} {\emph {\bibinfo {title} {Exciton Dynamics in
  Molecular Crystals and Aggregates: the Master Equation Approach}}}\ (\bibinfo
   {publisher} {Vol. 94 of Springer Tracts in Modern Physics},\ \bibinfo {year}
  {1982})\BibitemShut {NoStop}%
\bibitem [{\citenamefont {Kenkre}\ \emph {et~al.}(2008)\citenamefont {Kenkre},
  \citenamefont {Giuggioli},\ and\ \citenamefont
  {Kalay}}]{kenkre2008molecular}%
  \BibitemOpen
  \bibfield  {author} {\bibinfo {author} {\bibfnamefont {V.~M.}\ \bibnamefont
  {Kenkre}}, \bibinfo {author} {\bibfnamefont {L.}~\bibnamefont {Giuggioli}},\
  and\ \bibinfo {author} {\bibfnamefont {Z.}~\bibnamefont {Kalay}},\ }\bibfield
   {title} {\bibinfo {title} {Molecular motion in cell membranes: analytic
  study of fence-hindered random walks},\ }\href@noop {} {\bibfield  {journal}
  {\bibinfo  {journal} {Physical Review E}\ }\textbf {\bibinfo {volume} {77}},\
  \bibinfo {pages} {051907} (\bibinfo {year} {2008})}\BibitemShut {NoStop}%
\bibitem [{\citenamefont {Koszto{\l}owicz}(2001)}]{kosztolowicz2001random}%
  \BibitemOpen
  \bibfield  {author} {\bibinfo {author} {\bibfnamefont {T.}~\bibnamefont
  {Koszto{\l}owicz}},\ }\bibfield  {title} {\bibinfo {title} {Random walk in a
  discrete and continuous system with a thin membrane},\ }\href@noop {}
  {\bibfield  {journal} {\bibinfo  {journal} {Physica A: Statistical Mechanics
  and its Applications}\ }\textbf {\bibinfo {volume} {298}},\ \bibinfo {pages}
  {285} (\bibinfo {year} {2001})}\BibitemShut {NoStop}%
\bibitem [{\citenamefont {Powles}\ \emph {et~al.}(1992)\citenamefont {Powles},
  \citenamefont {Mallett}, \citenamefont {Rickayzen},\ and\ \citenamefont
  {Evans}}]{powles1992exact}%
  \BibitemOpen
  \bibfield  {author} {\bibinfo {author} {\bibfnamefont {J.~G.}\ \bibnamefont
  {Powles}}, \bibinfo {author} {\bibfnamefont {M.}~\bibnamefont {Mallett}},
  \bibinfo {author} {\bibfnamefont {G.}~\bibnamefont {Rickayzen}},\ and\
  \bibinfo {author} {\bibfnamefont {W.}~\bibnamefont {Evans}},\ }\bibfield
  {title} {\bibinfo {title} {Exact analytic solutions for diffusion impeded by
  an infinite array of partially permeable barriers},\ }\href@noop {}
  {\bibfield  {journal} {\bibinfo  {journal} {Proceedings of the Royal Society
  of London. Series A: Mathematical and Physical Sciences}\ }\textbf {\bibinfo
  {volume} {436}},\ \bibinfo {pages} {391} (\bibinfo {year}
  {1992})}\BibitemShut {NoStop}%
\bibitem [{\citenamefont {Dudko}\ \emph {et~al.}(2004)\citenamefont {Dudko},
  \citenamefont {Berezhkovskii},\ and\ \citenamefont
  {Weiss}}]{dudko2004diffusion}%
  \BibitemOpen
  \bibfield  {author} {\bibinfo {author} {\bibfnamefont {O.~K.}\ \bibnamefont
  {Dudko}}, \bibinfo {author} {\bibfnamefont {A.~M.}\ \bibnamefont
  {Berezhkovskii}},\ and\ \bibinfo {author} {\bibfnamefont {G.~H.}\
  \bibnamefont {Weiss}},\ }\bibfield  {title} {\bibinfo {title} {Diffusion in
  the presence of periodically spaced permeable membranes},\ }\href@noop {}
  {\bibfield  {journal} {\bibinfo  {journal} {The Journal of Chemical Physics}\
  }\textbf {\bibinfo {volume} {121}},\ \bibinfo {pages} {11283} (\bibinfo
  {year} {2004})}\BibitemShut {NoStop}%
\bibitem [{\citenamefont {Hahn}\ and\ \citenamefont
  {{\"O}zisik}(2012)}]{hahn2012heat}%
  \BibitemOpen
  \bibfield  {author} {\bibinfo {author} {\bibfnamefont {D.~W.}\ \bibnamefont
  {Hahn}}\ and\ \bibinfo {author} {\bibfnamefont {M.~N.}\ \bibnamefont
  {{\"O}zisik}},\ }\href@noop {} {\emph {\bibinfo {title} {Heat Conduction}}}\
  (\bibinfo  {publisher} {John Wiley \& Sons},\ \bibinfo {year}
  {2012})\BibitemShut {NoStop}%
\bibitem [{\citenamefont {Kay}\ \emph {et~al.}(2022)\citenamefont {Kay},
  \citenamefont {McKetterick},\ and\ \citenamefont
  {Giuggioli}}]{kay2022defect}%
  \BibitemOpen
  \bibfield  {author} {\bibinfo {author} {\bibfnamefont {T.}~\bibnamefont
  {Kay}}, \bibinfo {author} {\bibfnamefont {T.~J.}\ \bibnamefont
  {McKetterick}},\ and\ \bibinfo {author} {\bibfnamefont {L.}~\bibnamefont
  {Giuggioli}},\ }\bibfield  {title} {\bibinfo {title} {The defect technique
  for partially absorbing and reflecting boundaries: Application to the
  ornstein--uhlenbeck process},\ }\href@noop {} {\bibfield  {journal} {\bibinfo
   {journal} {International Journal of Modern Physics B}\ }\textbf {\bibinfo
  {volume} {36}},\ \bibinfo {pages} {2240011} (\bibinfo {year}
  {2022})}\BibitemShut {NoStop}%
\bibitem [{\citenamefont {Moutal}\ and\ \citenamefont
  {Grebenkov}(2019)}]{moutal2019diffusion}%
  \BibitemOpen
  \bibfield  {author} {\bibinfo {author} {\bibfnamefont {N.}~\bibnamefont
  {Moutal}}\ and\ \bibinfo {author} {\bibfnamefont {D.}~\bibnamefont
  {Grebenkov}},\ }\bibfield  {title} {\bibinfo {title} {Diffusion across
  semi-permeable barriers: spectral properties, efficient computation, and
  applications},\ }\href@noop {} {\bibfield  {journal} {\bibinfo  {journal}
  {Journal of Scientific Computing}\ }\textbf {\bibinfo {volume} {81}},\
  \bibinfo {pages} {1630} (\bibinfo {year} {2019})}\BibitemShut {NoStop}%
\bibitem [{\citenamefont {Novikov}\ \emph {et~al.}(2011)\citenamefont
  {Novikov}, \citenamefont {Fieremans}, \citenamefont {Jensen},\ and\
  \citenamefont {Helpern}}]{novikov2011random}%
  \BibitemOpen
  \bibfield  {author} {\bibinfo {author} {\bibfnamefont {D.~S.}\ \bibnamefont
  {Novikov}}, \bibinfo {author} {\bibfnamefont {E.}~\bibnamefont {Fieremans}},
  \bibinfo {author} {\bibfnamefont {J.~H.}\ \bibnamefont {Jensen}},\ and\
  \bibinfo {author} {\bibfnamefont {J.~A.}\ \bibnamefont {Helpern}},\
  }\bibfield  {title} {\bibinfo {title} {Random walks with barriers},\
  }\href@noop {} {\bibfield  {journal} {\bibinfo  {journal} {Nature Physics}\
  }\textbf {\bibinfo {volume} {7}},\ \bibinfo {pages} {508} (\bibinfo {year}
  {2011})}\BibitemShut {NoStop}%
\bibitem [{\citenamefont {Kenkre}(2021)}]{kenkre2021memory}%
  \BibitemOpen
  \bibfield  {author} {\bibinfo {author} {\bibfnamefont {V.~M.}\ \bibnamefont
  {Kenkre}},\ }\href@noop {} {\emph {\bibinfo {title} {Memory Functions,
  Projection Operators, and the Defect Technique: Some Tools of the Trade for
  the Condensed Matter Physicist}}},\ Vol.\ \bibinfo {volume} {982}\ (\bibinfo
  {publisher} {Springer Nature},\ \bibinfo {year} {2021})\BibitemShut {NoStop}%
\bibitem [{Note1()}]{Note1}%
  \BibitemOpen
  \bibinfo {note} {See Supplemental Material at [URL will be inserted by
  publisher] for mathematical details and derivations.}\BibitemShut {Stop}%
\bibitem [{\citenamefont {Tanner}(1978)}]{tanner1978transient}%
  \BibitemOpen
  \bibfield  {author} {\bibinfo {author} {\bibfnamefont {J.~E.}\ \bibnamefont
  {Tanner}},\ }\bibfield  {title} {\bibinfo {title} {Transient diffusion in a
  system partitioned by permeable barriers. application to nmr measurements
  with a pulsed field gradient},\ }\href@noop {} {\bibfield  {journal}
  {\bibinfo  {journal} {The Journal of Chemical Physics}\ }\textbf {\bibinfo
  {volume} {69}},\ \bibinfo {pages} {1748} (\bibinfo {year}
  {1978})}\BibitemShut {NoStop}%
\bibitem [{\citenamefont {Kosztolowicz}\ and\ \citenamefont
  {Mrowczynski}(2001)}]{kosztolowicz2001membrane}%
  \BibitemOpen
  \bibfield  {author} {\bibinfo {author} {\bibfnamefont {T.}~\bibnamefont
  {Kosztolowicz}}\ and\ \bibinfo {author} {\bibfnamefont {S.}~\bibnamefont
  {Mrowczynski}},\ }\bibfield  {title} {\bibinfo {title} {Membrane boundary
  condition},\ }\href@noop {} {\bibfield  {journal} {\bibinfo  {journal} {Acta
  Physica Polonica. Series B}\ }\textbf {\bibinfo {volume} {32}},\ \bibinfo
  {pages} {217} (\bibinfo {year} {2001})}\BibitemShut {NoStop}%
\bibitem [{Note2()}]{Note2}%
  \BibitemOpen
  \bibinfo {note} {$\protect \mathcal {J}_r(t)=f[\protect \mathcal
  {P}_r(t)-\protect \mathcal {P}_{r+1}(t)]$ is the spatially discrete analogue
  of the PBC.}\BibitemShut {Stop}%
\bibitem [{Note3()}]{Note3}%
  \BibitemOpen
  \bibinfo {note} {{When the propagator is translationally invariant, one may
  replace $\partial _{x_0}\setbox \z@ \hbox {\mathsurround \z@ $\textstyle
  G$}\mathaccent "0365{G}_0(x,\epsilon |x_b)$ by $\setbox \z@ \hbox
  {\mathsurround \z@ $\textstyle J$}\mathaccent "0365{J}_0(x,\epsilon |x_b)/D$
  and $\partial _{x_0}\setbox \z@ \hbox {\mathsurround \z@ $\textstyle
  J$}\mathaccent "0365{J}_0(x_b,\epsilon |x_b)$ by $-\partial _{x}\setbox \z@
  \hbox {\mathsurround \z@ $\textstyle J$}\mathaccent "0365{J}_0(x_b,\epsilon
  |x_b)$}}\BibitemShut {NoStop}%
\bibitem [{\citenamefont {Schuss}(2009)}]{schuss2009theory}%
  \BibitemOpen
  \bibfield  {author} {\bibinfo {author} {\bibfnamefont {Z.}~\bibnamefont
  {Schuss}},\ }\href@noop {} {\emph {\bibinfo {title} {Theory and Applications
  of Stochastic Processes: an Analytical Approach}}},\ Vol.\ \bibinfo {volume}
  {170}\ (\bibinfo  {publisher} {Springer Science \& Business Media},\ \bibinfo
  {year} {2009})\BibitemShut {NoStop}%
\bibitem [{\citenamefont {Collins}\ and\ \citenamefont
  {Kimball}(1949)}]{collins1949diffusion}%
  \BibitemOpen
  \bibfield  {author} {\bibinfo {author} {\bibfnamefont {F.~C.}\ \bibnamefont
  {Collins}}\ and\ \bibinfo {author} {\bibfnamefont {G.~E.}\ \bibnamefont
  {Kimball}},\ }\bibfield  {title} {\bibinfo {title} {Diffusion-controlled
  reaction rates},\ }\href@noop {} {\bibfield  {journal} {\bibinfo  {journal}
  {Journal of Colloid Science}\ }\textbf {\bibinfo {volume} {4}},\ \bibinfo
  {pages} {425} (\bibinfo {year} {1949})}\BibitemShut {NoStop}%
\bibitem [{\citenamefont {Waite}(1957)}]{waite1957theoretical}%
  \BibitemOpen
  \bibfield  {author} {\bibinfo {author} {\bibfnamefont {T.}~\bibnamefont
  {Waite}},\ }\bibfield  {title} {\bibinfo {title} {Theoretical treatment of
  the kinetics of diffusion-limited reactions},\ }\href@noop {} {\bibfield
  {journal} {\bibinfo  {journal} {Physical Review}\ }\textbf {\bibinfo {volume}
  {107}},\ \bibinfo {pages} {463} (\bibinfo {year} {1957})}\BibitemShut
  {NoStop}%
\bibitem [{\citenamefont {Weiss}(1994)}]{weiss1994aspects}%
  \BibitemOpen
  \bibfield  {author} {\bibinfo {author} {\bibfnamefont {G.}~\bibnamefont
  {Weiss}},\ }\href {https://books.google.co.uk/books?id=QRnvAAAAMAAJ} {\emph
  {\bibinfo {title} {Aspects and Applications of the Random Walk}}},\
  International Congress Series\ (\bibinfo  {publisher} {North-Holland},\
  \bibinfo {year} {1994})\BibitemShut {NoStop}%
\bibitem [{\citenamefont {Redner}(2001)}]{redner2001guide}%
  \BibitemOpen
  \bibfield  {author} {\bibinfo {author} {\bibfnamefont {S.}~\bibnamefont
  {Redner}},\ }\href@noop {} {\emph {\bibinfo {title} {A Guide to First-Passage
  Processes}}}\ (\bibinfo  {publisher} {Cambridge University Press},\ \bibinfo
  {year} {2001})\BibitemShut {NoStop}%
\bibitem [{\citenamefont {Feller}(2008)}]{feller2008introduction}%
  \BibitemOpen
  \bibfield  {author} {\bibinfo {author} {\bibfnamefont {W.}~\bibnamefont
  {Feller}},\ }\href@noop {} {\emph {\bibinfo {title} {An Introduction to
  Probability Theory and its Applications, Vol 2}}}\ (\bibinfo  {publisher}
  {John Wiley \& Sons},\ \bibinfo {year} {2008})\BibitemShut {NoStop}%
\bibitem [{\citenamefont {Abate}\ \emph {et~al.}(2000)\citenamefont {Abate},
  \citenamefont {Choudhury},\ and\ \citenamefont
  {Whitt}}]{abate2000introduction}%
  \BibitemOpen
  \bibfield  {author} {\bibinfo {author} {\bibfnamefont {J.}~\bibnamefont
  {Abate}}, \bibinfo {author} {\bibfnamefont {G.~L.}\ \bibnamefont
  {Choudhury}},\ and\ \bibinfo {author} {\bibfnamefont {W.}~\bibnamefont
  {Whitt}},\ }\bibfield  {title} {\bibinfo {title} {An introduction to
  numerical transform inversion and its application to probability models},\
  }in\ \href@noop {} {\emph {\bibinfo {booktitle} {Computational
  probability}}}\ (\bibinfo  {publisher} {Springer},\ \bibinfo {year} {2000})\
  pp.\ \bibinfo {pages} {257--323}\BibitemShut {NoStop}%
\bibitem [{\citenamefont {Kac}(1947)}]{kac1947notion}%
  \BibitemOpen
  \bibfield  {author} {\bibinfo {author} {\bibfnamefont {M.}~\bibnamefont
  {Kac}},\ }\bibfield  {title} {\bibinfo {title} {On the notion of recurrence
  in discrete stochastic processes},\ }\href@noop {} {\bibfield  {journal}
  {\bibinfo  {journal} {Bulletin of the American Mathematical Society}\
  }\textbf {\bibinfo {volume} {53}},\ \bibinfo {pages} {1002} (\bibinfo {year}
  {1947})}\BibitemShut {NoStop}%
\bibitem [{\citenamefont {L{\'e}vy}(1940)}]{levy1940certains}%
  \BibitemOpen
  \bibfield  {author} {\bibinfo {author} {\bibfnamefont {P.}~\bibnamefont
  {L{\'e}vy}},\ }\bibfield  {title} {\bibinfo {title} {Sur certains processus
  stochastiques homog{\`e}nes},\ }\href@noop {} {\bibfield  {journal} {\bibinfo
   {journal} {Compositio Mathematica}\ }\textbf {\bibinfo {volume} {7}},\
  \bibinfo {pages} {283} (\bibinfo {year} {1940})}\BibitemShut {NoStop}%
\bibitem [{\citenamefont {Majumdar}\ and\ \citenamefont
  {Comtet}(2002)}]{majumdar2002local}%
  \BibitemOpen
  \bibfield  {author} {\bibinfo {author} {\bibfnamefont {S.~N.}\ \bibnamefont
  {Majumdar}}\ and\ \bibinfo {author} {\bibfnamefont {A.}~\bibnamefont
  {Comtet}},\ }\bibfield  {title} {\bibinfo {title} {Local and occupation time
  of a particle diffusing in a random medium},\ }\href@noop {} {\bibfield
  {journal} {\bibinfo  {journal} {Physical Review Letters}\ }\textbf {\bibinfo
  {volume} {89}},\ \bibinfo {pages} {060601} (\bibinfo {year}
  {2002})}\BibitemShut {NoStop}%
\bibitem [{\citenamefont {Kac}(1949)}]{kac1949distributions}%
  \BibitemOpen
  \bibfield  {author} {\bibinfo {author} {\bibfnamefont {M.}~\bibnamefont
  {Kac}},\ }\bibfield  {title} {\bibinfo {title} {On distributions of certain
  wiener functionals},\ }\href@noop {} {\bibfield  {journal} {\bibinfo
  {journal} {Transactions of the American Mathematical Society}\ }\textbf
  {\bibinfo {volume} {65}},\ \bibinfo {pages} {1} (\bibinfo {year}
  {1949})}\BibitemShut {NoStop}%
\bibitem [{\citenamefont {Kac}(1951)}]{kac1951some}%
  \BibitemOpen
  \bibfield  {author} {\bibinfo {author} {\bibfnamefont {M.}~\bibnamefont
  {Kac}},\ }\bibfield  {title} {\bibinfo {title} {On some connections between
  probability theory and differential and integral equations},\ }in\ \href@noop
  {} {\emph {\bibinfo {booktitle} {Proceedings of the Second Berkeley Symposium
  on Mathematical Statistics and Probability}}}\ (\bibinfo {organization}
  {University of California Press},\ \bibinfo {year} {1951})\ pp.\ \bibinfo
  {pages} {189--215}\BibitemShut {NoStop}%
\bibitem [{\citenamefont {Majumdar}(2007)}]{majumdar2007brownian}%
  \BibitemOpen
  \bibfield  {author} {\bibinfo {author} {\bibfnamefont {S.~N.}\ \bibnamefont
  {Majumdar}},\ }\bibfield  {title} {\bibinfo {title} {Brownian functionals in
  physics and computer science},\ }in\ \href@noop {} {\emph {\bibinfo
  {booktitle} {The Legacy Of Albert Einstein: A Collection of Essays in
  Celebration of the Year of Physics}}}\ (\bibinfo  {publisher} {World
  Scientific},\ \bibinfo {year} {2007})\ pp.\ \bibinfo {pages}
  {93--129}\BibitemShut {NoStop}%
\bibitem [{\citenamefont {Tak{\'a}cs}(1995)}]{takacs1995local}%
  \BibitemOpen
  \bibfield  {author} {\bibinfo {author} {\bibfnamefont {L.}~\bibnamefont
  {Tak{\'a}cs}},\ }\bibfield  {title} {\bibinfo {title} {On the local time of
  the brownian motion},\ }\href@noop {} {\bibfield  {journal} {\bibinfo
  {journal} {The Annals of Applied Probability}\ ,\ \bibinfo {pages} {741}}
  (\bibinfo {year} {1995})}\BibitemShut {NoStop}%
\bibitem [{\citenamefont {Smoluchowski}(1915)}]{smoluchowski1915brownian}%
  \BibitemOpen
  \bibfield  {author} {\bibinfo {author} {\bibfnamefont {M.}~\bibnamefont
  {Smoluchowski}},\ }\bibfield  {title} {\bibinfo {title} {Brownian molecular
  movement under the action of external forces and its connection with the
  generalized diffusion equation},\ }\href@noop {} {\bibfield  {journal}
  {\bibinfo  {journal} {Annals of Physics (Leipzig)}\ }\textbf {\bibinfo
  {volume} {48}},\ \bibinfo {pages} {1103} (\bibinfo {year}
  {1915})}\BibitemShut {NoStop}%
\bibitem [{\citenamefont {Koszto{\l}owicz}\ and\ \citenamefont
  {Dutkiewicz}(2020)}]{kosztolowicz2020boundary}%
  \BibitemOpen
  \bibfield  {author} {\bibinfo {author} {\bibfnamefont {T.}~\bibnamefont
  {Koszto{\l}owicz}}\ and\ \bibinfo {author} {\bibfnamefont {A.}~\bibnamefont
  {Dutkiewicz}},\ }\bibfield  {title} {\bibinfo {title} {Boundary conditions at
  a thin membrane for normal diffusion, classical subdiffusion, and slow
  subdiffusion processes},\ }\href@noop {} {\bibfield  {journal} {\bibinfo
  {journal} {Mathematical Methods in the Applied Sciences}\ }\textbf {\bibinfo
  {volume} {43}},\ \bibinfo {pages} {10500} (\bibinfo {year}
  {2020})}\BibitemShut {NoStop}%
\end{thebibliography}%

\end{document}